\documentclass[dvipdfmx]{PoS}

\newcommand{\Slash}[1]{{\ooalign{\hfil/\hfil\crcr$#1$}}}

\title{
  Analysis of topological structure of the QCD vacuum 
  with overlap-Dirac operator eigenmode
}  

\ShortTitle{
  Analysis of topological structure of the QCD vacuum 
  with overlap-Dirac operator eigenmode}

\author{\speaker{T.~Iritani}$^a$, G.~Cossu$^a$, and S.~Hashimoto$^{ab}$ \\
\llap{$^a$}High Energy Accelerator Research Organization (KEK),
  Ibaraki 305-0801, Japan \\
  \llap{$^b$}School of High Energy Accelerator Science,
  The Graduate University for Advanced Studies (Sokendai),
  Ibaraki 305-0801, Japan \\
  E-mail: \email{iritani@post.kek.jp}
}

\abstract{
  Using the eigenmodes of the overlap-Dirac operator,
  we study the topological structure of the QCD vacuum.
  We investigate the space-time profile of the low-lying eigenmodes 
  and 
  their contribution to the vacuum action density and chiral condensate
  under the existence of static color sources.
  We demonstrate that the low-lying Dirac eigenmode
  shows the flux-tube structure, 
  which suggests the relevance to confinement.
  We also analyze the chiral condensate in the flux-tube.
  Chiral symmetry is partially restored 
  inside the flux, and the reduction of the condensate 
  is about 20\% at the center of the tube.
}
         
\FullConference{31st International Symposium on Lattice Field Theory LATTICE 2013\\
                 July 29 - August 3, 2013\\
                 Mainz, Germany}

\begin{document}

\section{Introduction}

The QCD vacuum has interesting topological objects
known as instantons/anti-instantons, 
which are (anti-)self-dual field configurations, 
and characterized by the topological charge.
Instantons are expected to contribute to chiral symmetry breaking,
and there are many studies using the instanton-based models \cite{Schafer:1996wv}.
The vacuum structure is modified by the existence color sources.
For instance, our typical understanding is that
a flux-tube is formed between quark and antiquark,
which is observed by the spatial distribution of action/energy density 
using lattice QCD \cite{Bali:1995,Haymaker:1996,Bissey:2007,Yamamoto:2009}.
This tube structure leads to the linear potential and confinement.
Thus, the vacuum structure itself is an interesting subject
for the dynamics of QCD.

The Dirac eigenmodes reflect non-perturbative properties of QCD.
For example, chiral symmetry breaking occurs due to 
an accumulation of near-zero Dirac eigenmodes \cite{BanksCasher},
and the number of zero-modes is related to the topological charge
of the background gauge fields \cite{AtiyahSinger}.
In lattice QCD, 
the overlap-Dirac operator is a useful formulation
to study the Dirac eigenmodes and vacuum structures,
since it keeps the exact chiral symmetry on lattice \cite{GinspargWilson},
and gives a proper definition of the topological properties.
The definition of the massless overlap-Dirac operator is given by
\begin{equation}
  D_{\rm ov}(0) = m_0 \left[ 1 + \gamma_5 \mathrm{sgn} \ H_W(-m_0) \right],
  \label{eq:overlap-Dirac-operator}
\end{equation}
using the hermitian Wilson-Dirac operator $H_W(-m_0) = \gamma_5 D_W(-m_0)$ 
and a sign function
\cite{Neuberger:1998}.

In this paper, we study the space-time profile of
the low-lying overlap-Dirac eigenmodes 
using the eigenmode decomposition of a field strength tensor $F_{\mu\nu}$ 
\cite{Gattringer:2002,Ilgenfritz:2007}.
In order to discuss the effect of valence quarks,
we also investigate a spatial distribution of the action density
and chiral condensate around the static color sources.

\section{Duality of the low-lying overlap-Dirac eigenmodes}

First, we introduce the Dirac eigenmode decomposition of
the field-strength tensor $F_{\mu\nu}$,
which is proposed by Gattringer in Ref.~\cite{Gattringer:2002}.
Let us consider a square of the Dirac operator $\Slash{D} \equiv \gamma_\mu D_\mu$,
\begin{equation}
  [\Slash{D}(x)]^2 = \sum_{\mu} D_\mu^2(x) 
  + \sum_{\mu<\nu} \gamma_\mu \gamma_\nu F_{\mu\nu}(x).
\end{equation}
By multiplying $\gamma_\mu\gamma_\nu$ and taking a trace
with respect to the Dirac indices
and by expanding in terms of the Dirac eigenmodes $\psi_\lambda$,
that satisfy $\Slash{D}\psi_\lambda = \lambda \psi_\lambda$,
the field-strength tensor
is expressed in terms of the Dirac operator as 
\begin{equation}
  F_{\mu\nu}(x) = - \frac{1}{4}\mathrm{tr}
  \left[ \gamma_\mu \gamma_\nu \Slash{D}^2(x) \right]
  = \sum_\lambda \lambda^2 f_{\mu\nu}(x)_{\lambda},
  \label{eq:f_decompose}
\end{equation}
with the field-strength tensor components
\begin{equation}
  f_{\mu\nu}(x)_\lambda \equiv 
  \frac{1}{2} \psi_\lambda^\dagger(x) \gamma_\mu \gamma_\nu \psi_\lambda(x).
\end{equation}

In this paper, 
we use the eigenmodes of the overlap-Dirac operator on
2+1-flavor dynamical overlap-fermion configurations
produced by the JLQCD Collaboration \cite{JLQCD}.
The lattice is of $16^3 \times 48$ at $\beta = 2.3$,
which corresponds to $a^{-1} = 1.759(10)$~GeV.
The quark masses are $m_{ud} = 0.015$ and $m_s = 0.080$,
and the topological charge is fixed at $Q = 0$.

\begin{figure}
  \centering
  \includegraphics[width=0.47\textwidth,clip]{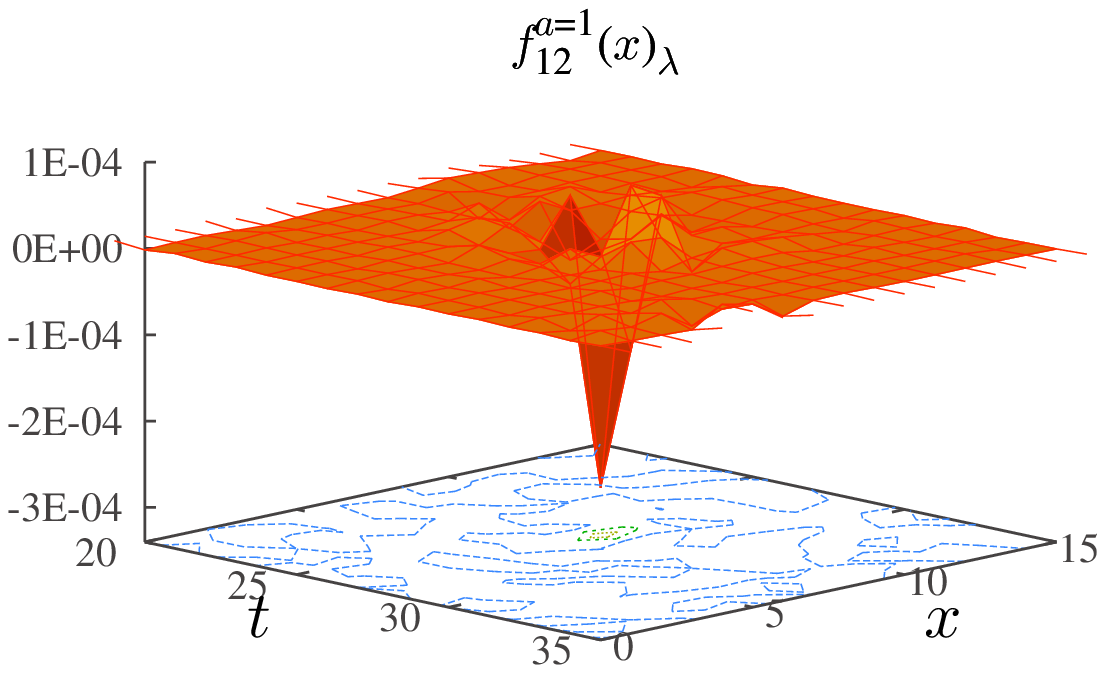}
  \includegraphics[width=0.47\textwidth,clip]{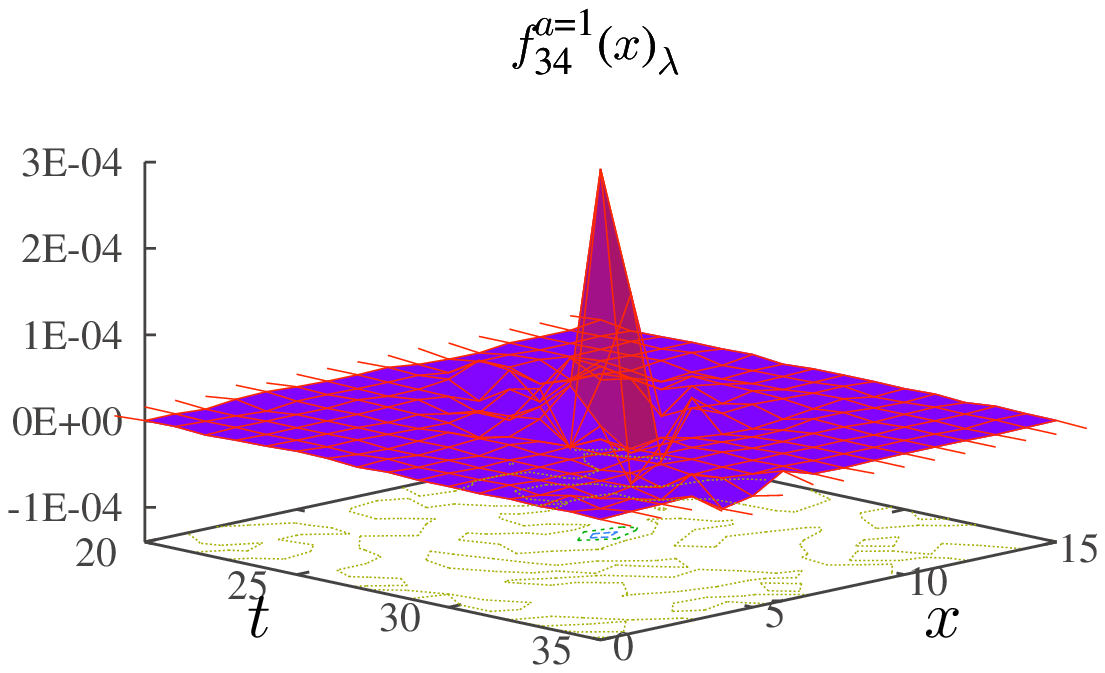}
  \caption{
    \label{fig:fmunu-component}
    The lowest Dirac eigenmode of field strength tensor 
    $f_{\mu\nu}(x)_\lambda \equiv 2i\mathrm{Tr}[f_{\mu\nu}(x)_\lambda T^a]$
    on $xt$-plane at fixed $y$ and $z$ in $16^3 \times 48$ lattice.
    (a) $f_{12}^{a=1}(x)_\lambda$. 
    (b) $\widetilde{f^{a=1}_{12}}(x)_\lambda = f_{34}^{a=1}(x)_\lambda$.
  }
\end{figure}

Figure \ref{fig:fmunu-component} shows 
a snapshot of the lowest (non-zero) eigenmode 
of the field strength tensor $f_{12}^{a}(x)_\lambda
\equiv 2i\mathrm{Tr}[f_{12}(x)_\lambda T^a]$ and its dual component
$\widetilde{f_{12}^{a}}(x)_\lambda = f_{34}^{a}(x)_\lambda$ 
on a slice of four-dimensional lattice volume.
As shown in these figures, $f_{12}^{a}(x)_\lambda$ has a negative peak,
while $f_{34}^{a}(x)_\lambda$ shows a positive peak 
at the same location with almost the same magnitude,
which means that the existence of anti-self-dual lump.
Note that $f_{\mu\nu}(x)_\lambda$ has several (anti-)self-dual peaks
in four-dimensional space-time, and Fig.~\ref{fig:fmunu-component} shows one of the peaks.

We can also construct the action density and the topological charge density
in terms of the Dirac eigenmodes \cite{Gattringer:2002}.
Using the field strength tensor $f_{\mu\nu}(x)_\lambda$,
the low-mode truncated action density $\rho^{(N)}$
and the topological charge density $q^{(N)}$  
are given by
\begin{equation}
  \rho^{(N)}(x) \equiv 
  \sum_{i,j}^{N}
  \frac{\lambda_i^2 \lambda_j^2}{2}
  f_{\mu\nu}^a(x)_i f_{\mu\nu}^a(x)_j,
  \quad
  q^{(N)}(x) \equiv \sum_{i,j}^{N}
  \frac{\lambda_i^2 \lambda_j^2}{2}
  f_{\mu\nu}^a(x)_i \widetilde{f_{\mu\nu}^a}(x)_j,
  \label{eq:action_dens_low_mode}
\end{equation}
with $N$ the number of the eigenmodes, respectively.

\begin{figure}
  \centering
  \includegraphics[width=0.47\textwidth,clip]{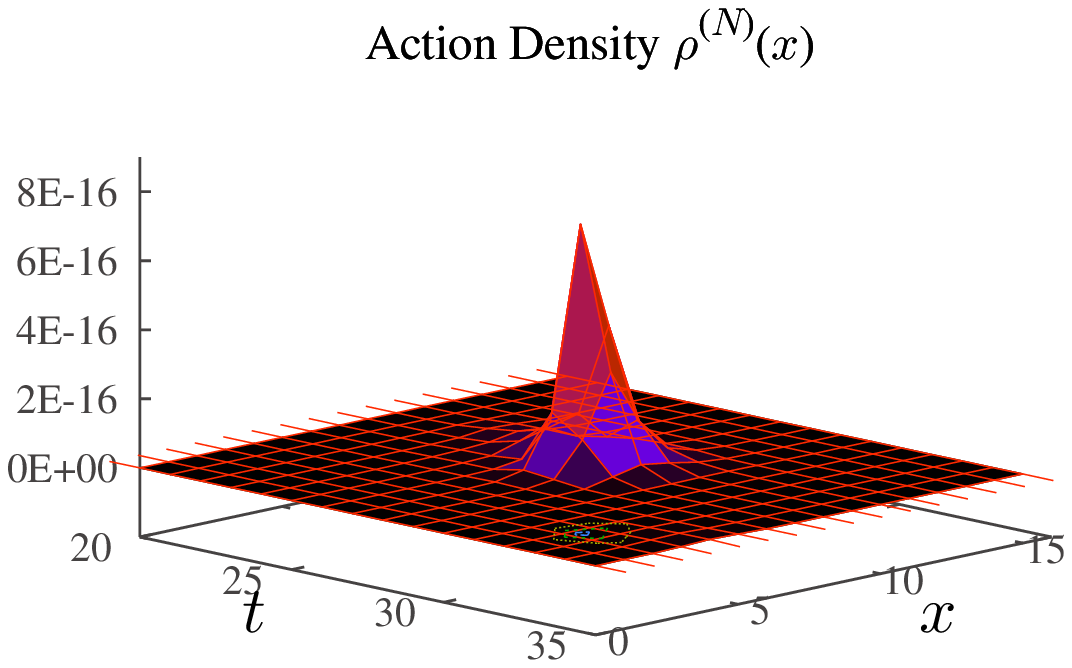}
  \includegraphics[width=0.47\textwidth,clip]{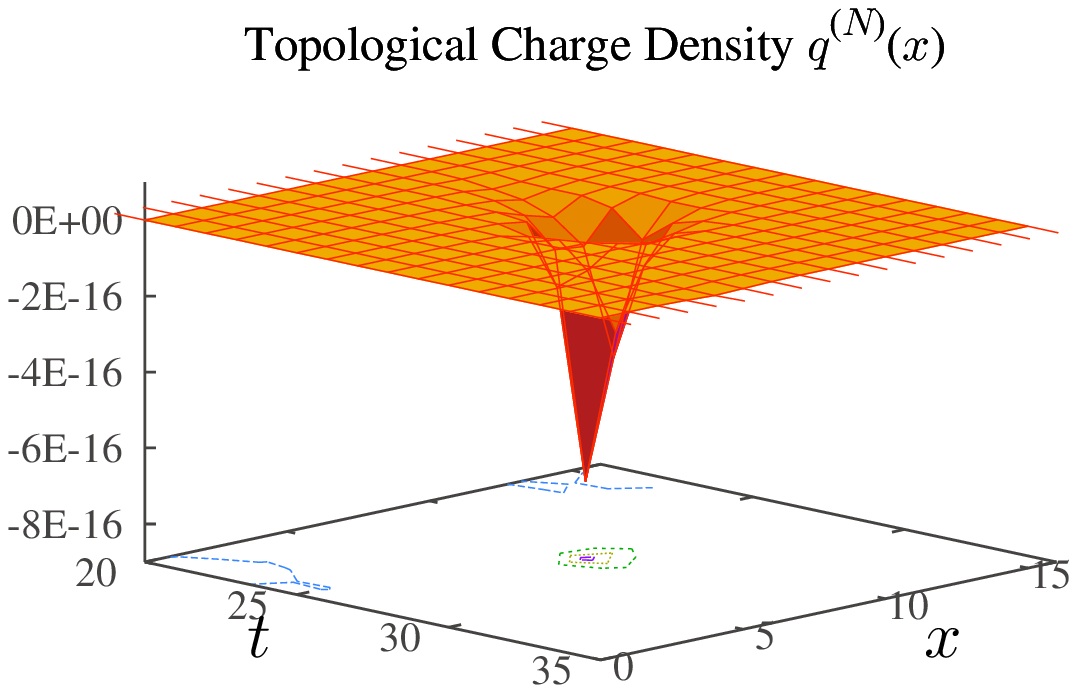}
  \caption{
    \label{fig:action-topolo-sample}
    The lowest mode component of the action density $\rho^{(N)}(x)$ and
    the topological charge density $q^{(N)}(x)$
    on $xt$-plane at fixed $y$ and $z$.
    (a) The action density $\rho^{(N)}(x)$.
    (b) The topological charge density $\rho^{(N)}(x)$.
  }
\end{figure}

Figure \ref{fig:action-topolo-sample}
shows the lowest (non-zero) eigenmode component of
the action density and the topological charge density 
on the same plane in Fig.~\ref{fig:fmunu-component}.
These figures show the existence of an anti-instanton.
We also checked that other low-lying eigenmodes show
(anti-)self-dual peaks similar to 
those shown in Figs.~\ref{fig:fmunu-component} and \ref{fig:action-topolo-sample}.

\section{Flux-tube formation by the low-lying Dirac eigenmodes}

Next, we discuss a distribution of the action density
under the existence of static color sources.
In particular, we consider
a spatial distribution around the Wilson loop.
Since chiral symmetry breaking originates from the accumulation
of the near-zero eigenmodes \cite{BanksCasher},
it is interesting to investigate their role in the confining force among color sources,
which may give some insight to
the relation between chiral symmetry breaking and confinement
\cite{Gongyo:2012}.

\begin{figure}
  \centering
  \includegraphics[width=0.55\textwidth,clip]{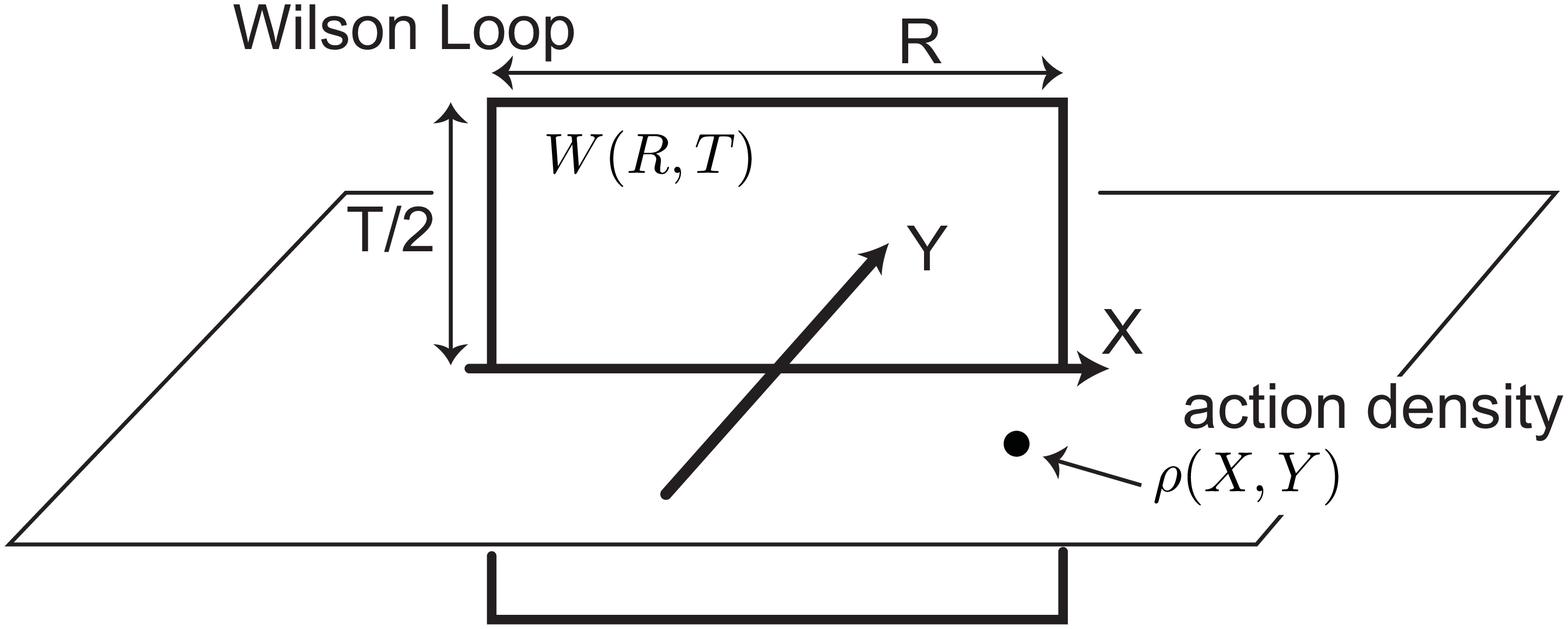}
  \caption{ \label{fig:flux-tube-measurement}
  A schematic picture of the flux-tube measurement on the lattice.
  Quark and antiquark are located at $(R/2,0)$ and $(-R/2,0)$.
}
\end{figure}

We look at a spatial distribution of the action density $\rho(x)$ around 
the Wilson loop $W(R,T)$ as
\begin{equation}
  \langle \rho(x) \rangle_W \equiv \frac{\langle \rho(x)W(R,T) \rangle}{\langle W(R,T) \rangle}
  - \langle \rho \rangle,
  \label{eq:fluxTubeMeasurement}
\end{equation}
where $R \times T$ is the size of the Wilson loop 
\cite{Bali:1995,Haymaker:1996,Bissey:2007,Yamamoto:2009}.
Figure \ref{fig:flux-tube-measurement} shows
a schematic picture of the measurement on the lattice.
Here, the origin of a coordinate is the center of the Wilson loop $W(R,T)$,
and quark and antiquark are located at $(R/2,0)$ and $(-R/2,0)$.
We use the low-mode truncated action density $\rho^{(N)}(x)$
as $\rho(x)$ in Eq.~(\ref{eq:fluxTubeMeasurement}).

\begin{figure}
  \centering
  \includegraphics[width=0.49\textwidth,clip]{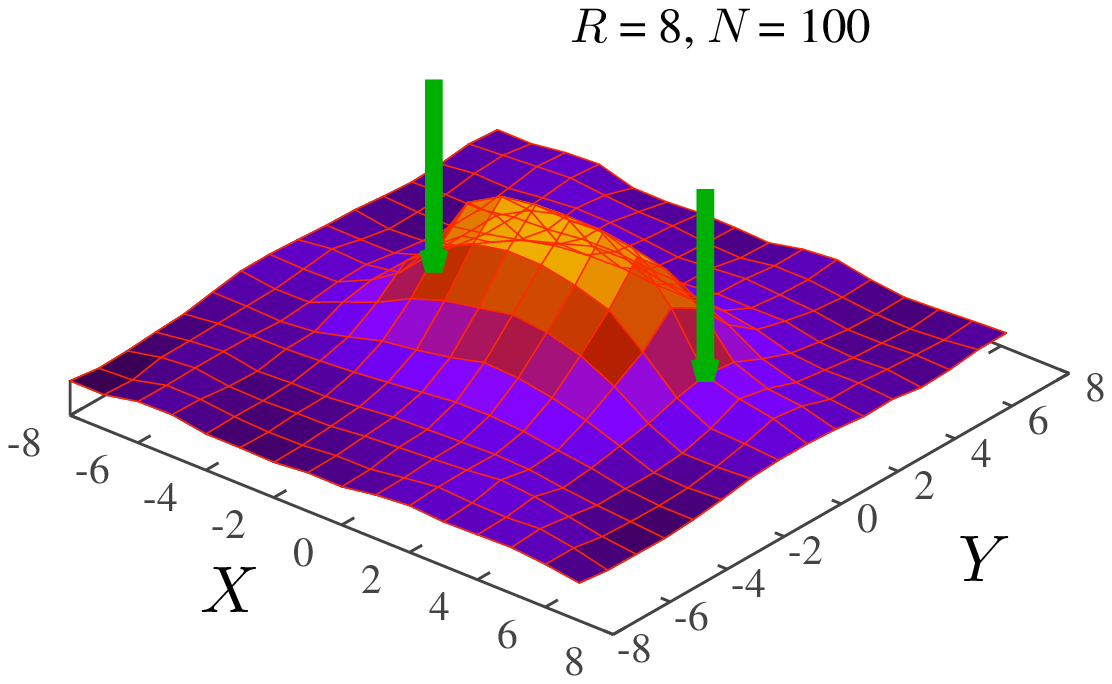}
  \includegraphics[width=0.49\textwidth,clip]{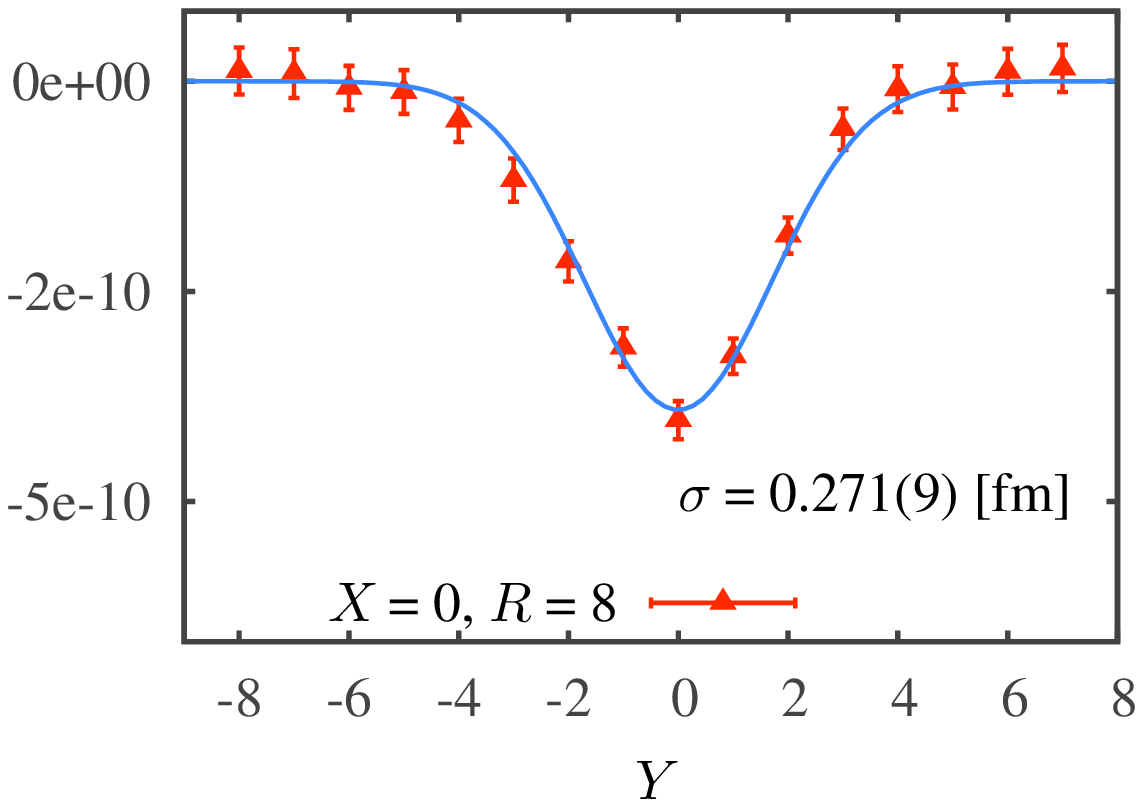}
  \caption{ \label{fig:flux-tube}
  (a) The spatial distribution of the truncated action density
    $(-)\langle \rho(x) \rangle_W^{(N)}$ around the Wilson loop $W(R,T)$
    using 100 low-lying eigenmodes with $R = 8$.
    The vertical bars indicate positions of (anti-)quarks.
  (b) The cross-section of $\langle \rho(x) \rangle_W^{(N)}$ 
    along $Y$-axis in the center of flux-tube $X = 0$,
    a curve denotes fit result using a gaussian function
    $e^{-Y^2/\sigma^2}$.
  }
\end{figure}

Figure \ref{fig:flux-tube} (a) shows $\langle \rho(x) \rangle_W^{(N)}$ with
a quark separation $R = 8$.
In order to improve the signal, we adopt the APE smearing for the spatial link-variables.
We measure the Wilson loop with $T = 4$, where the ground state becomes dominant,
and the number of configurations is 50.
We truncate the sum at $N = 100$, which corresponds to 
$\lambda \lesssim 300$~MeV.
As shown in Fig.~\ref{fig:flux-tube} (a), 
we observe a tube-like structure, which indeed stretches towards 
the $X$-direction as $R$ increases.
We note that the number of eigenmodes is only 100 out of
$16^3 \times 48 \times 3 \times 4 = 2,359,296$ modes.
This figure indicates that these low-lying modes
reflect the confining nature of the QCD vacuum.

Figure \ref{fig:flux-tube} (b) is the cross-section of the tube
in the center of the Wilson loop.
We estimate a thickness of the flux-tube by fitting with a gaussian form $e^{-Y^2/\sigma^2}$,
which describes our data very well.
A typical value of the thickness is about 0.6~fm,
which is consistent with the value of the previous lattice QCD studies 
\cite{Bali:1995,Haymaker:1996,Bissey:2007,Yamamoto:2009}.

\section{Partial restoration of chiral symmetry inside the flux-tube}

Next, we analyze a change of the chiral condensate around the static color sources.
Using the overlap-Dirac eigenfunction $\psi_\lambda(x)$,
a local chiral condensate is defined as
\begin{equation}
  \bar{q}q(x) = - \sum_\lambda \frac{\psi_\lambda^\dagger(x)\psi_\lambda(x)}{m_q
    + (1 + \frac{m_q}{2m_0})\lambda},
  \label{eq:localChiralCondensate}
\end{equation}
where $\lambda$ denotes eigenvalues of the massless overlap-Dirac operator
in Eq.~(\ref{eq:overlap-Dirac-operator}),
and $m_q$ a quark mass.
The chiral condensate $\langle \bar{q}q \rangle$ is given by 
an expectation value of $\bar{q}q(x)$.

We measure the local chiral condensate $\bar{q}q(x)$ around 
the static color sources as
\begin{equation}
    \langle \bar{q}q(x) \rangle_W 
    \equiv \frac{\langle\bar{q}q(x) W(R,T)\rangle}{\langle W(R,T)\rangle}
    - \langle \bar{q}q \rangle,
  \label{eq:diffLocalChiral}
\end{equation}
with the Wilson loop $W(R,T)$.
Here, we use a low-mode truncated condensate 
$\bar{q}q^{(N)}(x)$ by truncating the sum in Eq.~(\ref{eq:localChiralCondensate}).

\begin{figure}
  \centering
  \includegraphics[width=0.49\textwidth,clip]{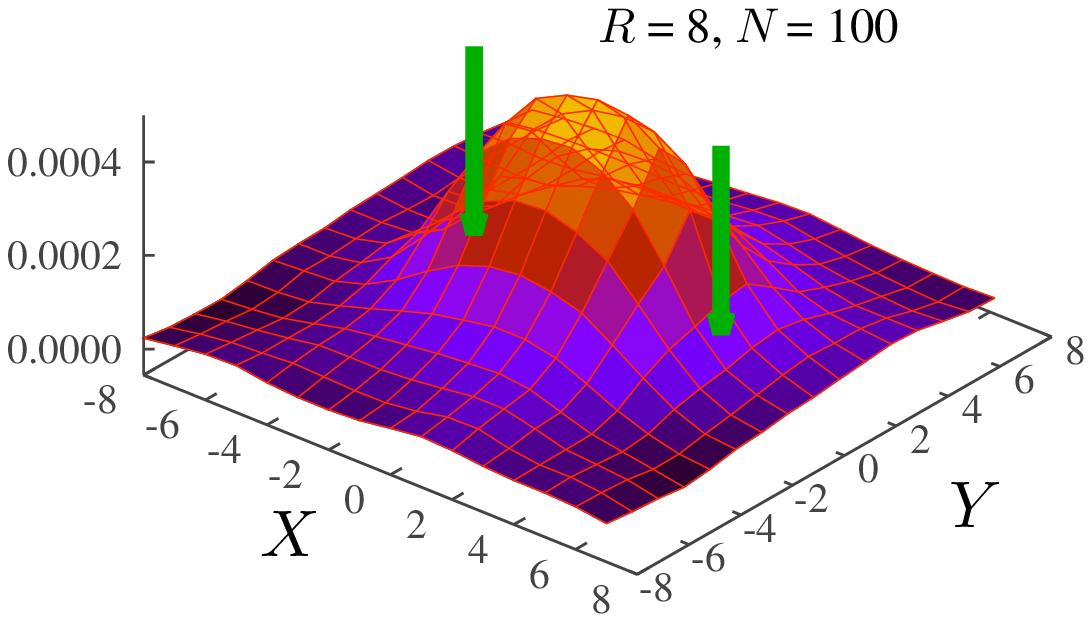}
  \includegraphics[width=0.49\textwidth,clip]{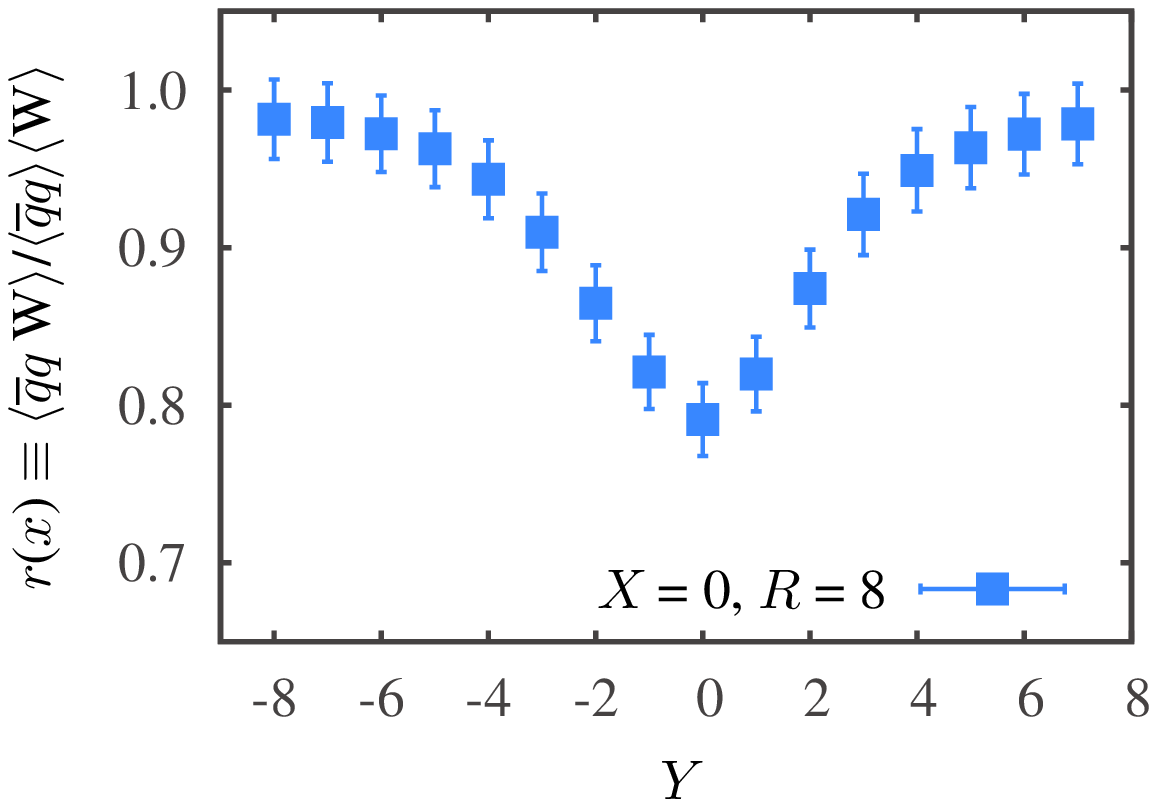}
  \caption{
    \label{fig:localChiral}
    (a) The spatial distribution of the local chiral condensate 
    $\langle \bar{q}q(x) \rangle_{W}$ around the Wilson loop $W(R,T)$ with $R = 8$.
    The vertical bars denote the position of (anti-)quark.
    (b) The cross-section of the ratio $r(X=0,Y)$ 
        in the center of quark-antiquark $X = 0$.}
\end{figure}

Figure \ref{fig:localChiral} (a)
shows $\langle \bar{q}q(x) \rangle_W^{(N)}$ with a number of eigenmodes $N = 100$, 
and the quark mass $m_q = 0.015$. 
The configuration of the Wilson loop 
is the same as that in Fig.~\ref{fig:flux-tube}.
We note that the change of the local chiral condensate becomes positive
between the color sources.
It suggests that the magnitude of chiral condensate is reduced,
since the chiral condensate has a negative expectation value in the vacuum.

In order to discuss a quantitative change of the chiral condensate,
we need to renormalize the divergent operator $\bar{q}q$.
Here, we use the mode truncation as a regularization method
\cite{Noaki:2009xi}.
Due to the exact chiral symmetry of the overlap-Dirac operator,
the power divergence is parameterized as
\begin{equation}
  \langle \bar{q}q \rangle^{(N)} = 
  \langle \bar{q}q^{\rm (subt)} \rangle  
  + {c_1^{(N)}m_q/a^2} + c_2^{(N)}m_q^3
  \label{eq:subtractChiral}
\end{equation}
where $\langle \bar{q}q^{\rm (subt)} \rangle$
is the subtracted condensate, which is finite up to logarithmic divergence \cite{Noaki:2009xi}.
For a ratio
\begin{equation}
  r(x) \equiv \frac{\langle \bar{q}q^{\rm (subt)}(x)W(R,T) \rangle}
  {\langle \bar{q}q^{\rm (subt)} \rangle \langle W(R,T)\rangle},
  \label{eq:reductionRatio}
\end{equation}
the remaining divergence also cancels.
Figure \ref{fig:localChiral} (b) 
shows the ratio $r(x)$ along $Y$-axis in the center of the tube ($X=0$).
It means that chiral symmetry is partially restored inside the flux.
The reduction of the local chiral condensate is estimated about 20\% at the core of the flux.

\section{Summary and Discussion}

In this work, we study the non-trivial vacuum structure of QCD
in terms of the overlap-Dirac eigenmodes.
We demonstrate that the low-lying overlap-Dirac eigenmodes
show instanton-like structure
using the eigenmodes decomposition of the field strength tensor $F_{\mu\nu}(x)$.

We also have analyzed a spatial distribution of the action density around the Wilson loop,
and discussed the flux-tube formation as seen by the Dirac eigenmodes.
We have shown that the low-lying modes indeed form 
the tube structure between the color sources.
This result does not necessarily mean that
the low-lying eigenmodes are essential degrees of freedom 
for both confinement and chiral symmetry breaking.
In fact, it is reported that confinement still remains even without low-lying eigenmodes
\cite{Gongyo:2012,Lang:2011}.
These results may suggest that there is no specific momentum scales for confinement 
when we observe through the Dirac eigenmodes \cite{Gongyo:2012}.
Further studies are needed for understanding of the relation 
between confinement and chiral symmetry breaking.

Finally, we have investigated the local chiral condensate around the Wilson loop.
We have shown that the magnitude of the condensate is reduced inside the flux.
It suggests that chiral symmetry is partially restored between the static color sources,
which may be interpreted as a bag-model-like state.
Similar results are also shown 
by measuring the chiral condensate around the Polyakov loop
in lattice QCD \cite{Faber:1993},
and estimating the flux-tube effect from Nambu-Jona-Lasinio model \cite{Suganuma:1990nn}.
It is also interesting to study 
the color charge effects on various kinds of operators,
such as the quark number density and the topological charge.

\section*{Acknowledgements}
  The lattice QCD calculations have been done on SR16000 at 
  High Energy Accelerator Research Organization (KEK)
  under a support of its Large Scale Simulation Program (No. 12-05).
  This work is supported in part by the Grant-in-Aid of the Japanese
  Ministry of Education (No. 21674002),
  and the SPIRE (Strategic Program for Innovative REsearch) Field5 project.


\begin{thebibliography}{99}
    \bibitem{Schafer:1996wv}
    T.~Sch\"afer and E.~V.~Shuryak,
    \emph{Rev.\ Mod.\ Phys.} {\bf 70} (1998) 323 [hep-ph/9610451].

  \bibitem{Bali:1995}
    G.~S.~Bali, K.~Schilling and C.~Schlichter,
    \emph{Phys.\ Rev.} {\bf D51} (1995) 5165 [hep-lat/9409005].

  \bibitem{Haymaker:1996}
    R.~W.~Haymaker, V.~Singh, Y.~-C.~Peng and J.~Wosiek,
    \emph{Phys.\ Rev.} {\bf D53} (1996) 389 [hep-lat/9406021].

  \bibitem{Bissey:2007}
    F.~Bissey, F-G.~Cao, A.~R.~Kitson, A.~I.~Signal,
    D.~B.~Leinweber, B.~G.~Lasscock and A.~G.~Williams,
    \emph{Phys.\ Rev.} {\bf D76} (2007) 114512, [hep-lat/0606016].

  \bibitem{Yamamoto:2009}
    A.~Yamamoto,
    \emph{Phys.\ Lett.\ B} {\bf 688} (2010) 345 [arXiv:0906.2618 [hep-lat]].

  \bibitem{BanksCasher}
    T.~Banks and A.~Casher, \emph{Nucl. Phys.} {\bf B169} (1980) 103.

  \bibitem{AtiyahSinger}
    M.~F.~Atiyah and I.~M.~Singer, \emph{Ann. Math.} {\bf 87} (1968) 484.

  \bibitem{GinspargWilson}
    P.~H.~Ginsparg and K.~G.~Wilson,
    \emph{Phys. Rev.} {\bf D25} (1982) 2649.

  \bibitem{Neuberger:1998}
    H.~Neuberger, \emph{Phys. Lett. B} {\bf 417} (1998) 141
    [hep-lat/9707022];
    \emph{ibid.} \textbf{427} (1998) 353 [hep-lat/9801031].

  \bibitem{Gattringer:2002} 
    C.~Gattringer, 
    \emph{Phys. Rev. Lett.} {\bf 88} (2002) 221601 [hep-lat/020220].

  \bibitem{Ilgenfritz:2007} 
    E.~-M.~Ilgenfritz, K.~Koller, Y.~Koma, G.~Schierholz, T.~Streuer, 
    and V.~Weinberg, \emph{Phys.\ Rev.} {\bf D76} (2007) 034506 
    [arXiv:0705.0018 [hep-lat]];
    E.~-M.~Ilgenfritz, D.~Leinweber, P.~Moran, K.~Koller, G.~Schierholz, 
    and V.~Weinberg, \emph{Phys.\ Rev.} {\bf D77} (2008) 074502 
    [Erratum-ibid.\ D {\bf 77} (2008) 099902],
    [arXiv:0801.1725 [hep-lat]], and references therein.

  \bibitem{JLQCD}
    S.~Aoki et al. (JLQCD and TWQCD Collaborations), 
    \emph{Prog. Theor. Exp. Phys.} {\bf 2012} (2012) 01A106.

  \bibitem{Gongyo:2012}
  S.~Gongyo, T.~Iritani and H.~Suganuma, \emph{Phys.\ Rev.} {\bf D86} (2012) 034510
  [arXiv:1202.4130 [hep-lat]].  T.~Iritani and H.~Suganuma, [arXiv:1305.4049 [hep-lat]].

  \bibitem{Noaki:2009xi} 
    J.~Noaki, T.~W.~Chiu, H.~Fukaya, S.~Hashimoto, H.~Matsufuru, 
    T.~Onogi, E.~Shintani and N.~Yamada,
    \emph{Phys.\ Rev.} {\bf D81} (2010) 034502
    [arXiv:0907.2751 [hep-lat]].

  \bibitem{Lang:2011}
    C.B.~Lang and M.~Schr\"ock, \emph{Phys. Rev.} {\bf D84} (2011) 087704
    [arXiv:1107.5195 [hep-lat]];
    L.Ya.~Glozman, C.B.~Lang, and M.~Schr\"ock, \emph{ibid.} {\bf D86} (2012) 014507
    [arXiv:1205.4887 [hep-lat]].

  \bibitem{Faber:1993}
    M.~Faber, M.~Schaler, and H.~Gausterer,
    \emph{Phys. Lett. B} {\bf 317} (1993) 409;
    W.~Sakuler, W.~Burger, M.~Faber, H.~Markum, M.~Muller, P.~De Forcrand,
    A.~Nakamura and I.~O.~Stamatescu,
    \emph{Phys.\ Lett.\ B} {\bf 276} (1992) 155;
    S.~Thurner, M.~Feurstein, H.~Markum, and W.~Sakuler,
    \emph{Phys.\ Rev.} {\bf D54} (1996) 3457;
    K.~H\"ubner, 
    \pos{PoS (Lattice 2007) 193} [arXiv:0709.1467 [hep-lat]].

  \bibitem{Suganuma:1990nn} 
    H.~Suganuma and T.~Tatsumi,
    \emph{Annals Phys.}\  {\bf 208} (1991) 470;
    \emph{Phys.\ Lett.\ B} {\bf 269} (1991) 371;
    \emph{Prog. Theor. Phys.} {\bf 90} (1993) 379.

\end{thebibliography}
\end{document}